# The role of substrate bias and nitrogen doping on the structural evolution and local elastic modulus of diamond-like carbon films

S. R. Polaki,*,† K. Ganesan,* S. K. Srivastava, M. Kamruddin,† and A. K. Tyagi,†

Materials Science Group, Indira Gandhi Centre for Atomic Research, Kalpakkam 603 102, India.
†*Homi Bhabha National Institute, Anushaktinagar, Mumbai - 400 094, India.*

## Abstract

Diamond-like carbon (DLC) films are synthesized on Si using plasma enhanced chemical vapor deposition. The role of substrate bias and nitrogen doping on the structural evolution and local elastic modulus of DLC films are systematically investigated. Raman spectroscopic studies reveal that the amount of graphitic C=C $sp^2$ bonding increases with substrate bias and nitrogen doping. The density and hydrogen concentration in the films are found to vary from 0.7 to 2.2 g/cm$^3$ and 16 to 38 atomic %, respectively, depending upon the substrate bias and nitrogen concentration in the DLC films. Atomic force acoustic microscopic (AFAM) analysis shows a direct correlation between local elastic modulus and structural properties estimated by Raman spectroscopy, Rutherford back scattering and elastic recoil detection analysis. AFAM analysis further confirms the evolution of soft second phases at high substrate biases ($\geq$-150V) in undoped DLC films. Further, N doping leads to formation of such soft second phases in DLC films even at lower substrate bias of -100 V. The AFAM studies provide a direct microscopic evidence for the "sub-implantation growth model" which predicts the formation of graphitic second phases in DLC matrix at high substrate biases.

Key words: Diamond like carbon films, Raman Spectroscopy, Acoustic Force Atomic microscopy, Elastic modulus

# 1. Introduction

Diamond like carbon (DLC) films attracted intense research interest due to its excellent physical and chemical properties such as high hardness, ultra-low frictional coefficient, high wear resistance, chemical inertness and bio-compatibility [1]. Owing to these exotic characteristics, DLC films are potential candidate for a wide range of applications which include as protective coatings on cutting tools, memory storage devices, bio-medical devices, anti-reflection coating on infrared optical windows and Si solar cells [1-6]. The DLC films are also extensively used for tribological applications due to their exceptionally low frictional co-efficient [7, 8]. In addition, DLC films also find applications in micro-electromechanical systems as a structural material which is used to fabricate high frequency resonators and comb-drives for communication and sensing applications [5, 6]. Despite having several reports on synthesis, deposition of ultra thin DLC films with homogenous elastic and mechanical properties is still challenging [3, 9-11]. In DLC films, the elastic or mechanical characteristics mainly depend on the amount of C-C $sp^3$ bonding. But, the amount C-C $sp^3$ bonding is mainly influenced by ion energy which is about equal to substrate bias and also other growth parameters such as deposition temperature and chemical composition (doping) [1, 9-11]. The ion energy of C radicals can be easily tuned by controlling substrate bias. It is established that the DLC films are found to attain maximum C-C $sp^3$ bonding at an optimum ion energy of about 100 eV, which can be attained by applying substrate bias of -100V [1]. However, the applied substrate bias also introduces compressive residual stress into the films [1, 12]. Such residual stress can be reduced in DLC films by doping with metal/nonmetal elements [12]. Among various dopants, nitrogen is widely studied dopant in DLC films. Nitrogen doping helps not only to reduce the stress but also improves the toughness of the DLC films [12]. But, high substrate bias and nitrogen doping also increase the amount of C=C $sp^2$ bonding which leads to structural in-homogeneity in the DLC films. Raman spectroscopy is the most commonly used tool to probe the structural characteristics which provides the ratio of $sp^2/sp^3$ bonding and size of graphitic clusters in the DLC films [13, 14]. However, variations in elastic properties at nano-scale resolution cannot be probed by Raman spectroscopy.

Evaluation of elastic properties of ultra thin films at nanometer-scale resolution demand specialized characterization tools due to complexity in the film-substrate elastic coupling. Novel tools like nano-indentation, surface Brillouin light scattering (SBS), surface wave ultrasonic techniques and atomic force acoustic microscope (AFAM) are most commonly used to obtain elastic properties of ultra thin films [15-17]. However, nano-indentation and ultrasonic based techniques require slightly higher film thickness to avoid the influence of substrate [16]. Though SBS is used to obtain modulus of ultra thin films (less than 10nm), but it requires prior knowledge on film density which has to be measured independently [16]. The AFAM is one of the powerful techniques to obtain quantitative elastic properties of thin films [17-23]. Also, AFAM allows mapping elastic modulus of materials at nanometer-scale lateral resolution which makes the technique to be unique over all other techniques. Moreover, the applied static loads in AFAM measurements are higher than the adhesion forces but still low enough to cause any plastic deformation in the film [18]. Thus, AFAM is advantageous, in particular to characterize ultrathin films and nanostructures. Due to these advantages, many reports do exist on the estimation of elastic modulus of DLC films [19-22]. For example, the elastic modulus of DLC films are estimated very accurately for thickness even less than 10 nm using sensitivity enhanced AFAM having a concentrated-mass cantilever with flat tip [23]. The presence of soft glassy carbon in the DLC matrix is detected by mapping elastic modulus which is varying from 140 to 180 GPa in the films grown by pulsed laser deposition [22]. However, a systematic AFAM studies on series of DLC films with varying structural properties and their direct correlation with elastic moduli are still scarce.

Historically, the growth mechanism of DLC films are explained in terms of sub-implantation model which predicts evolution of graphitic second phases under high energy sub-surface ion implantation during growth [24-27]. Further, the sub-implantation growth model is explained extensively using molecular dynamics simulations [27-29]. But, the experimental verification is mostly performed using Raman spectroscopy [13, 14]. Although a few reports are available on co-existing soft and hard phases in DLC films estimated by AFAM, evolution of elastically soft second phases as a function of ion energy is not reported in literature. Here, we make a systematic study on a series of pure and nitrogen doped DLC films that are grown under different ion energy by suitably varying substrate bias. The structural phase evaluation is thoroughly carried out using Raman spectroscopy, Rutherford back scattering (RBS) and elastic recoil detection analysis (ERDA) and AFAM. Further, a quantitative local elastic modulus mapping of these films is also obtained using AFAM. We observed a direct correlation between structural properties and local elastic modulus of the DLC films. Based on the structural evolution and their mechanical properties, a possible growth mechanism of DLC films under different substrate bias and nitrogen doping is discussed. Further, this study provides a direct microscopic evidence for evolution of soft second phases in DLC matrix under high substrate bias and nitrogen concentration through AFAM analysis.

## 2. Experimental Details

### 2.1 Growth

DLC films were grown on Si (100) substrates using electron cyclotron resonance plasma enhanced chemical vapor deposition. The details on experimental set up and deposition chamber were published elsewhere [7, 8]. The base and operating vacuum of the deposition chamber were $5 \times 10^{-6}$ and $2 \times 10^{-3}$ mbar, respectively. The DLC films were grown using ultra high purity feed stock gases $CH_4$ and Ar at a flow rate ratio of 6:9 respectively. The first set of DLC films were grown at different substrate biases viz. -50, -100, -150 and -200V using a pulsed DC power supply at a frequency of 50 kHz and these samples are labeled as $B_1$, $B_2$, $B_3$ and $B_4$, respectively. The second set of DLC films were grown with $CH_4$ + Ar : nitrogen at flow rate ratio of 15:0.5, 15:1 and 15:2.5 and also these films are labeled as $B_{2-0.5}$, $B_{2-1}$ and $B_{2-2.5}$, respectively. An optimized substrate bias of -100V at 50 kHz were used for nitrogen doping and all other growth parameters were kept constant. The growth duration was 120 minutes and the growth was carried out at room temperature.

## 2.2 Characterization

The degree of structural disorder in the DLC films were analyzed using micro-Raman spectrometer (inVia, M/s Renishaw, UK) with excitation laser wavelength of 514 nm. RBS and ERDA were carried out in 1.7 MV Tandetron particle accelerator for obtaining the density and hydrogen content in the films, respectively. For RBS studies $^4He^{2+}$ion beam with energy of 3.8 MeV and backscattered by 165°, was used. A similar ion beam with energy of 2.8 MeV, incident at 73°, scattered at 30°, was used for ERDA measurements. A mylar foil of thickness 10 μm was placed in front of a detector to stop scattered atoms heavier than hydrogen. Hydrogen passivated Si (100) was used as reference material for H quantification. SIMNRA software was used to fit RBS and ERDA spectra. Both RBS and ERDA spectra were fitted in an iterative way in such a way that a common composition and thickness was obtained. Surface chemical analysis and estimation of nitrogen concentration in DLC films were carried out by X-ray photoelectron spectroscopy (XPS, M/s SPECS, Germany) system equipped with a monochromatic Al Kα source (1486.7 eV).

## 2.3 Atomic force acoustic microscopy

A multi mode scanning probe microscope (Ntegra Aura, M/s. NT-MDT, The Netherlands) was used to carry out AFAM measurements. Silicon cantilevers having force constant of 45 N/m and length of 250 microns were used for the study. The elastic modulus of the film is calculated using standard model and the principle of AFAM analysis is briefly discussed here. In AFAM, the sample and AFM probe are set to vibrate together at an ultrasound frequency with an external piezoelectric transducer. By analyzing the contact mechanics of tip-sample vibration, the elastic properties of unknown materials are estimated. Here, the tip-sample contact stiffness depends on the Young's modulus of the sample and the tip, the load exerted by the tip and the geometry of the surface. Thus, AFAM can be used to determine the Young's modulus (E) of unknown materials from the contact stiffness when the cantilever properties are known [30].

According to Hertzian contact model for elastic deformation, contact stiffness (k*) of a hemispherical indenter with contact radius R on a flat surface with a force $F_c$ can be expressed as [30, 31]

$$k^* = \sqrt[3]{6RF_C E^{*2}} \quad \text{..................... (1)}$$

where E* is the reduced modulus, and can be expressed as

$$\frac{1}{E^*} = \frac{1-\nu_t}{E_t^*} + \frac{1-\nu_s}{E_s^*} \quad \text{......................(2)}$$

for an isotropic medium. Here, $E_t$, $\upsilon_t$, $E_s$, and $\upsilon_s$ are modulus and Poisson ratio of tip and sample, respectively. According to equation 1, the radius of curvature of the tip is an important factor in estimation of elastic modulus but it is not constant during measurements due to wearable nature of tip. In addition, it is also difficult to measure exact radius of the tip repeatedly. The uncertainty in the measurement of R leads to large error in calculated elastic modulus. Such errors can be avoided by performing a calibration test on a reference sample with known value of elastic modulus [30, 31]. The cantilever vibration mechanics can be solved numerically by several methods to calculate the elastic modulus of unknown materials. According to a simplified point mass model, in which elastic clamp is fixed at one end and the other end is with a point mass the contact stiffness of the coupled system can be written as [31]

$$f_s = \sqrt{k_s^* + k / m} \quad \cdots\cdots\cdots\cdots\cdots(3)$$

where $f_s$ is the contact resonance frequency (CRF) on the sample, $k_s^*$ is the effective tip-sample contact stiffness, $k$ is the spring constant of the cantilever and $m$ is the equivalent mass in point mass model. AFAM measurements on a sample with known contact stiffness ($k_s^*$) and CRF ($f_s$), allows to calculate the contact stiffness on a unknown sample by [32]

$$k_s^* = k_{ref}^* \left( \frac{f_s^2 - f_0^2}{f_{ref}^2 - f_0^2} \right) \quad \cdots\cdots\cdots\cdots(4)$$

where $k_s^*$ and $k_{ref}^*$ are the contact stiffness of the unknown and reference material, respectively.

The $f_s$, $f_{ref}$ and $f_0$ are the resonance frequency of the cantilever in contact with sample, reference material and free vibrations in air, respectively. The relation between elastic modulus and contact stiffness is represented as [33]

$$E_s^* = E_{ref}^* \left( \frac{k_s^*}{k_{ref}^*} \right)^n \quad \text{.......................... (5)}$$

Where n = 1 or 3/2 for flat or spherical shape probe, respectively. Si (100) is used as reference sample and its modulus value of 170 GPa is taken for calculation.

## 3. Results

### 3.1 Raman spectroscopy

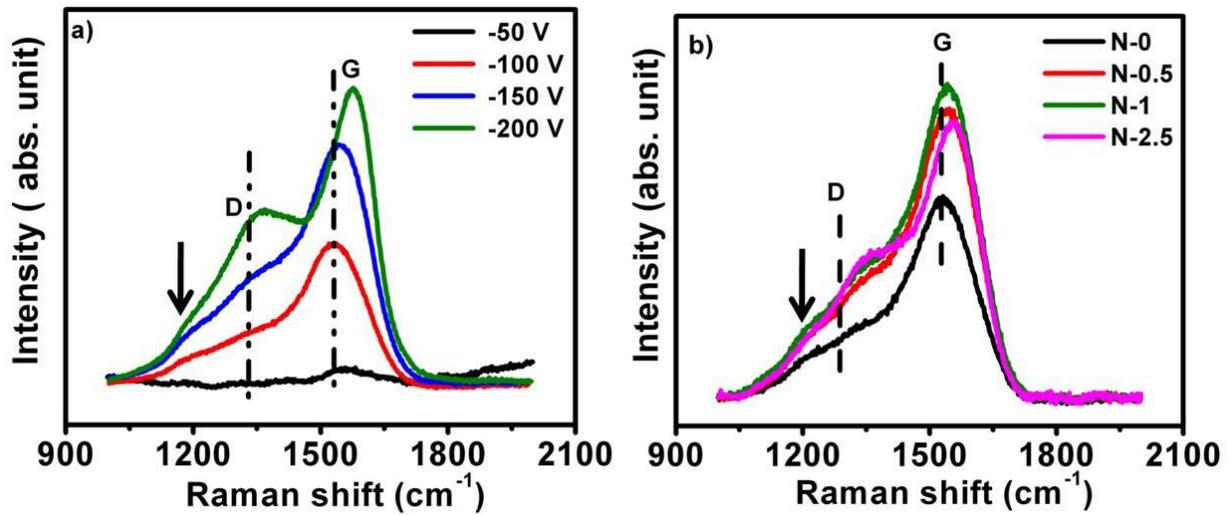

Fig.1 Raman spectra of the DLC films grown at different a) substrate biases and b) nitrogen flow rates

Figure 1 shows the Raman spectra of the DLC films grown at different substrate bias and nitrogen flow rates. These Raman spectra exhibit a typical disordered carbon film character with two prominent broad bands at around 1350 and 1580 cm$^{-1}$corresponding to D and G bands, respectively. The D peak is a disorder induced band which arises due to breathing mode of $sp^2$ atoms in rings and it is assigned to zone boundary phonon of A$_{1g}$ mode. Whereas the G band occurs from E$_{2g}$ symmetry mode and is due to bond stretching of pairs of C=C$sp^2$ atoms in both rings and chains [1]. A small hump at around 1200 cm$^{-1}$ indicates the existence of the trans-polyactelene structures [7, 8]. The Raman spectrum corresponding to the film deposited at -50 V bias ($B_1$) do not exhibit D and G bands clearly because of high luminescence background which is common for DLC films with high hydrogen content [1]. With increase in substrate bias, the intensity of D band increases and results in splitting of D and G bands at -200 V. The Raman spectra are deconvoluted into four Gaussian profiles and the extracted Raman parameters such as $I_D/I_G$ ratio, position and FWHM of G peak are given in table 1. As shown in table 1, the evolution of D band is reflected with increase in $I_D/I_G$ ratio with substrate bias. Further, the position and FWHM of G band increases and decreases with bias, respectively. These signatures indicate that the applied substrate bias helps to evolve more $sp^2$-rich clusters in DLC films [1, 13]. As the $sp^2$-rich cluster size increases with bias, the disorder in the films decreases and results in decrease of FWHM of graphitic G band [13]. The shift in G peak position from 1530 cm$^{-1}$ to higher wave numbers also reconfirms the increase in size of $sp^2$-rich clusters. The nitrogen doping in DLC films do not significantly alter Raman spectra as depicted in Fig.1b. The vibrational frequencies of solid carbon nitride are expected to lay very close to the modes of the analogous unsaturated CN molecules, which fall in the range of 1000 – 1800 cm$^{-1}$. Thus, it is extremely difficult to assess from D and G bands whether an aromatic ring contains nitrogen or not [13]. The increase in $I_D/I_G$ ratio and G peak position indicate the increase in $sp^2$-rich clusters with nitrogen doping. Thus, the nitrogen doping enhances the clustering of $sp^2$ sites even at lower substrate bias [31, 32].

Table 1.The growth parameters, thickness and Raman analysis of the diamond-like carbon films

| Sample | Bias (V) | CH4:Ar:N2 (sccm) | Thickness (nm) | Position(G) (cm$^{-1}$) | FWHM(G) (cm$^{-1}$) | $I_D/I_G$ ratio | N concen. (at % ) |
|---|---|---|---|---|---|---|---|
| **B1** | **-50** | 6:9:0 | 850 | 1531.0. | 176 | 0.36 | - |
| **B2** | **-100** | 6:9:0 | 475 | 1542.0 | 159 | 0.40 | - |
| **B3** | **-150** | 6:9:0 | 290 | 1555.0 | 147 | 0.55 | - |
| **B4** | **-200** | 6:9:0 | 240 | 1580.0 | 114 | 0.76 | - |
| **B2-0.5** | **-100** | 6:9:0.5 | 330 | 1548.0 | 144 | 0.47 | < 1.0 |
| **B2-1.0** | **-100** | 6:9:1.0 | 290 | 1551.0 | 137 | 0.54 | 5.7 |
| **B2-2.5** | **-100** | 6:9:2.5 | 200 | 1560.0 | 133 | 0.60 | 9.5 |

To confirm doping and quantification of nitrogen in DLC films, XPS measurements are carried out. XPS analyses reveal the nitrogen concentrations of 5.4, and 9.5 % for the samples B$_{2-1.0}$ and B$_{2-2.5,}$ respectively. The nitrogen concentration could not be quantified for the sample with B$_{2-0.5}$ since the N1s peak is below the XPS detection limit. C1s and N1s spectra for the DLC films grown at different substrate biases and nitrogen concentration are given in supplementary Fig. S1. Further, XPS analysis were detailed elsewhere [34]. We note, the estimation of nitrogen concentration of less than 1 atomic % in DLC films is not very accurate by XPS. As can be seen from table 1, the I$_D$/I$_G$ ratio, position and FWHM of G band vary with nitrogen concentration similar to that observed in films grown with increasing bias.

### 3.2 Film density and hydrogen concentration

Figure 2 shows the variation of film density and hydrogen concentration in the DLC films as a function of substrate bias and N flowrate as calculated from RBS and ERDA. At substrate bias of -50V, the film has the lowest density of 0.7 g/cm$^3$. Then, the film density increases to 2.1 g/cm³ at -100V and thereafter decreases with increasing bias. Whereas, in the case of N doped DLC, the film density monotonically decreases with doping as shown in Fig. 2b. The H concentration in DLC films also decreases monotonically with bias and N doping as depicted in Fig.2a and 2b, respectively. These observations clearly indicate that structural transformation takes place in the films with bias and N doping. Raman analyses also affirm such phase transformation in the films with increase in $sp^2$-rich cluster size with bias and N doping. Such variations in the density and H concentration of the DLC films are attributed to the sub-implantation process during growth [1] and it is elaborated in the discussion part.

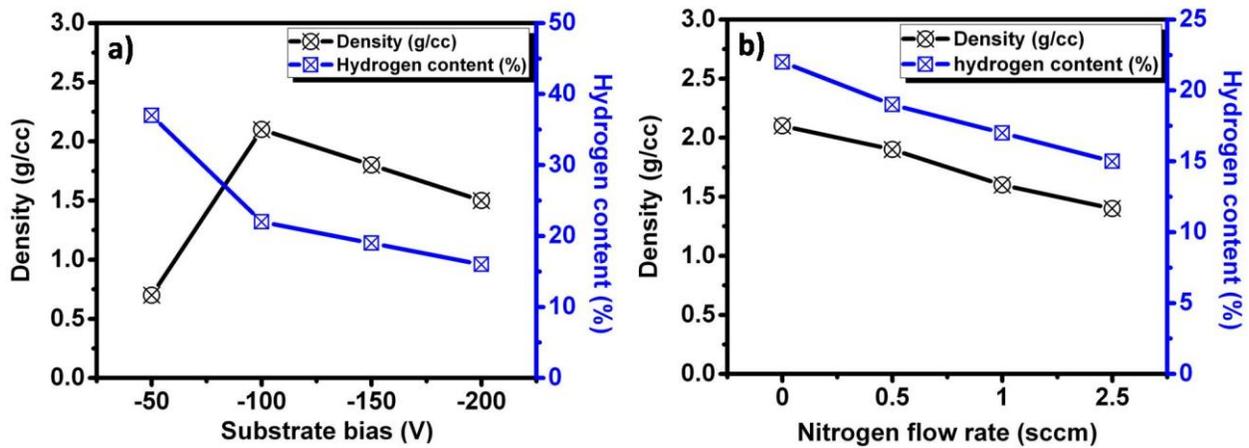

Fig. 2 Variation of density and hydrogen content in the DLC films grown under different conditions a) substrate bias and b) nitrogen flow rate.

**3.3 Atomic Force Acoustic Microscopy (AFAM) analysis**

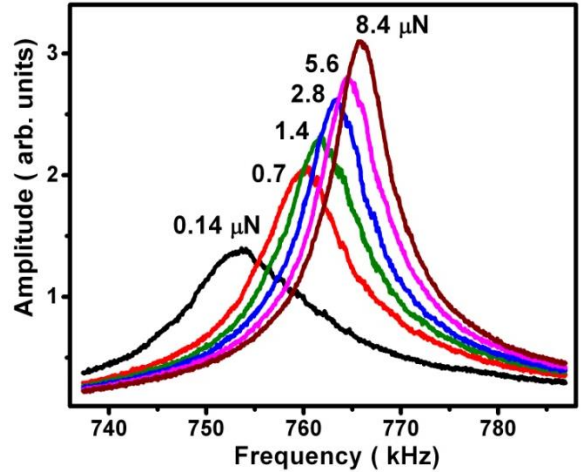

Fig. 3 The variation of contact resonance frequency as a function of load on the diamond-like carbon film grown at -100V.

The CRF of the tip − sample assembly is an important parameter in estimating the elastic modulus of the material. Several cursory CRF measurements are performed with different loads at various locations on these DLC films. A typical CRF spectra acquired at different loads are given in Fig. 3 for the sample $B_2$. While the load on tip increases, the peak position and amplitude of CRF increase but FWHM of these spectra decreases. These changes in CRF spectra indicate the good interaction of ultrasound waves with tip of the AFM cantilever. After having preliminary knowledge on CRF spectra, an optimized load of 2.8 µN is chosen for CRF mapping. Also, the AFAM measurements are repeated on each samples at different locations and scan areas. The results are found to consistent with each measurements.

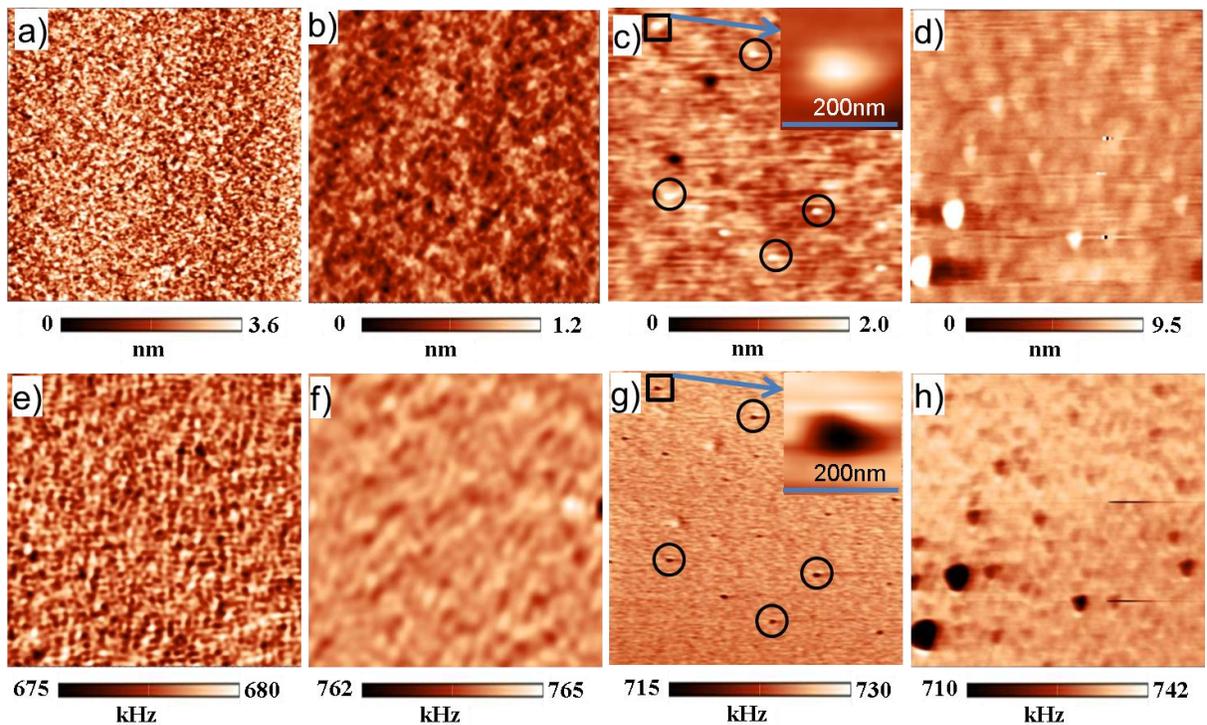

Fig. 4 Simultaneously recorded topography and contact resonance frequency mapping of DLC films grown at different substrate biases of (a,e) -50, (b,f) -100, (c,g) -150, and (d,h) -200V, respectively. The circle marks on Fig. 4c & 4g indicate the evolution of graphitic second phases in the DLC matrix. A magnified part of a cluster is also given as inset in Fig. 4c & 4g. The scan area is 5 x 5 µm² in all the images. The scale bar for topography and contact resonance frequency is given just below the respective images.

Figure 4(a-d) and (e-h) shows the simultaneously recorded topography and AFAM CRF mapping of DLC films grown at different substrate biases, respectively. The topography of the samples $B_1$ and $B_2$ are found to be extremely smooth with root mean square roughness of 0.5 and 0.14 nm, respectively. The sample $B_3$ exhibits isolated clusters of about 2 and 50 nm in height and lateral diameter respectively, over an extremely smooth surface. The rms roughness of the $B_3$ film is found to be 0.16 nm. But, the sample $B_4$ exhibits an inhomogeneous surface with large rms roughness of 4.5 nm. The corresponding CRF mapping of the samples $B_1$ and $B_2$ are also very smooth as shown in Fig.4 (e& f), respectively and have a small variation from 675 to 680

and 762 to 765 kHz, respectively. Similarly, the sample B₃ also exhibits a smooth CRF mapping but several isolated low CRF regimes are observed over the homogeneous background. In addition, a very careful topography and atomic force acoustic microscopy measurements are repeated on different locations and length scales in order to rule out any possible artifacts that could be seen as spikes in Fig. 4c & 4g. However, the observed results clearly confirm that they are indeed second phase and a magnified image of a typical second phase is given as inset in Fig. 4c & 4g. These low CRF regimes are found to coincide with topography mapping corresponding to the nano-clusters. Thus, we can infer from the CRF mapping that the nano-clustered regimes have lower elastic modulus than the matrix. Figure 4h shows the CRF mapping of the sample B₄ which varies from 710 to 742 kHz. As similar to the topography, the CRF mapping is also highly inhomogeneous with mixture of low and high frequency regimes indicating soft and hard modulus phases.

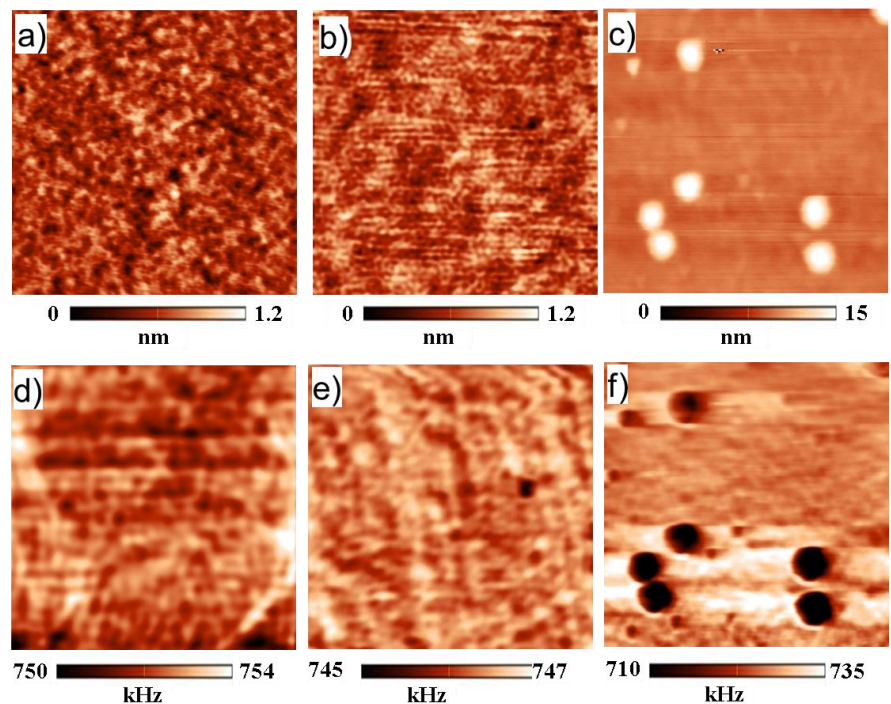

Fig. 5 Simultaneously recorded topography and contact resonance frequency mapping of DLC films grown at different nitrogen flow rate of (a,d) 0.5 (b,e) 1 and (c,f) 2.5 sccm, respectively. The scan area is 5 x 5 $\mu m^2$ in all the images. The scale bar for topography and contact resonance frequency is given just below the respective images.

Figure 5 shows the simultaneously recorded topography and AFAM mapping of DLC films with N doping. As discussed earlier, the DLC film with lower N concentration ($B_{2-0.5}$) also exhibits a homogeneous topography and CRF mapping as shown in Fig.5a and 5d, respectively. While the N concentration in DLC film is increased, the topography does not vary significantly but the CRFs are found to be much smaller than the sample $B_{2-0.5}$ as shown in Fig. 5b and 5e. The reduction in CRFs can be attributed to the increase in $sp^2$ bonding in the DLC matrix with nitrogen doping. A further increase in N concentration in the film increases the $sp^2$-rich cluster size in lateral dimension and evolves as second phases as given in Fig. 5c. The corresponding AFAM mapping of these samples also reveal that these second phase particles have lower elastic modulus than the matrix. Thus, the increase in nitrogen concentration in DLC leads to the formation of second phases, as similar to the phenomena observed with increase in applied bias. Further, a statistical analysis is carried out to extract the mean CRF value for every sample. Then, the CRF is converted into elastic modulus based on the equations 3 and 4. In order to avoid tip induced error in elastic modulus due to wear, the tip was always scanned over reference silicon before and after AFAM measurements on actual DLC films.

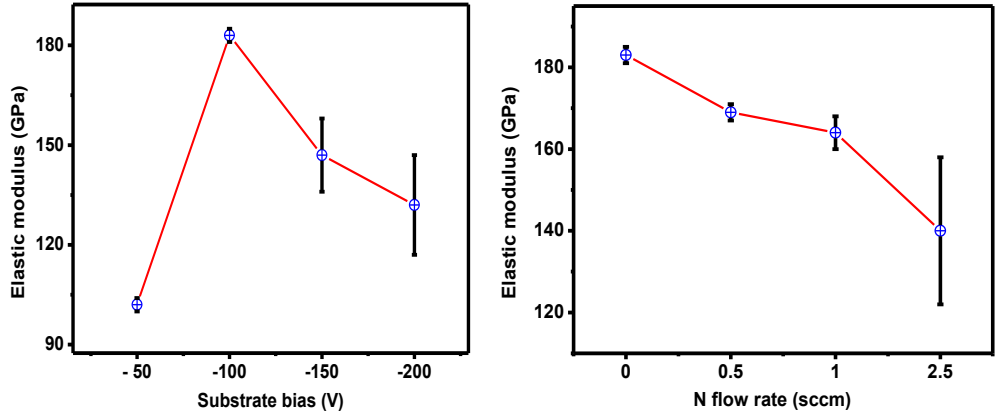

Fig. 6 The calculated elastic moduli of the DLC films grown at different a) bias and b) nitrogen flow rate. The solid line in the plots is guide to eye.

Figure 6 shows the variation in elastic modulus of the DLC films as a function bias and nitrogen flow rate. The elastic modulus of sample $B_1$ is found to be 102 GPa which is much lower than the reference Si. As the bias increases to -100V, E of the sample $B_2$ enhanced to 183 GPa which is higher than the reference Si. However, further increase in substrate bias to -150 and -200V, lead to decrease in E to 147 and 132 GPa, respectively. E reduces monotonically from 183 to 140 GPa with increasing N concentration in DLC as shown in Fig.6b. The reduced E can be attributed to the increase in C-C $sp^2$ bonding in the films with bias and N concentration as evidence from Raman measurements. The inter relation between the growth mechanism, chemical bonding, density and H concentration with elastic modulus of DLC films is discussed in next section.

## 4. Discussion

The observed results shall be summarized as follows. The Raman studies confirm that the $sp^2$ content increases with bias as well as nitrogen doping. The RBS and ERDA measurements also reveal that the density of the film initially increases when the substrate bias increases from -50 to -100V and then density starts to decrease monotonically with bias. Similarly, the density monotonically decreases with N incorporation into the DLC films that are grown at a bias of -100V. The density estimation from RBS and ERDA provide a direct support to Raman observations of increase in $sp^2$ concentration in DLC films. The increase in $sp^2$ content and variation of density in these DLC films as a function of substrate bias and N concentration can be explained by invoking an established growth mechanism called sub-implantation model [24, 25].

According to the sub-implantation model, the ion energy of C radicals decides the type of C-C bonds in DLC films. The ion energy mainly depends on the applied substrate bias during growth. At low substrate biases, the C ion energy is lower than the penetration threshold energy ($E_p$) (~30 eV) and hence, C ions just stay on surface and relax to minimum energy state by forming $sp^2$ bond with glassy structure. However, H is also incorporated in the growing film as the energy required to form C-H bonding is minimum [1, 37]. The H incorporation in the glassy carbon favors C-H $sp^3$ bonding but the film density remains very low [1, 25]. At -50V bias, the C ions gain enough energy to penetrate to the sub-surface and occupies an interstitial site. During sub-implantation growth process the locally altered bonding around the penetrated atoms reforms itself to become bulk bonding of the film with appropriate hybridization i.e. the atomic hybridization will adjust easily to change according to the local density under high energy ion bombardments [24]. Hence, the DLC film becomes $sp^2$ dominant if the local density is low and $sp^3$ dominant if the local density is high [25]. At -100V bias, a large amount of C ions enter sub-surface of the growing film since the C ions have sufficient energy. Since larger C ions occupy interstitial sites and C-C $sp^3$ bonding is energetically favored which in turn increase the film density. In the meantime, H is also removed from the growing film since the C-H bonds get easily broken at high energy ion bombardments.

At substrate bias greater than -100V, C ions gain much higher energy and it penetrates deeper into the film. The excess energy of C ions get dissipated in different processes viz. a) to penetrate into the subsurface, b) atomic displacements along the penetration track and c) release of phonons which results in thermal spikes at sub-pico second time scale [26]. This thermal spike allows the atoms to diffuse back to the surface and relax the density locally to stabilize with $sp^2$ bonding [1, 27, 28]. The size of the $sp^2$-rich clusters further increases with bias. Thus, the film density is large at substrate bias of -100 V and it reduces with increase in bias as shown in Fig. 2. Further, thickness of film also reduces with the increase in bias as shown in the table 1. The reduction in thickness is attributed to the re-sputtering of surface atoms by high energy C ions bombardments. In addition, the increase in ion energy also causes preferential sputtering of the lighter hydrogen atoms from the DLC films [1]. Thus, reduction in hydrogen content also occurs with increase in substrate bias [1, 37].

In case of N doped DLC films, the sub-implantation process takes place as discussed above with high efficiency since the films are grown at optimum ion energy of about 100eV. Here, the N bonds with carbon in either σ ($sp^3$) or π ($sp^2$) configuration [13]. However, the C-N $sp^2$ bonding is more favorable than the former [38]. Further, N incorporation encourages the formation of C=C $sp^2$ bonding because cross linking H is removed from the C-H $sp^3$ network [23]. Once local C-N $sp^2$ bond is formed, it further grows as $sp^2$-rich cluster in larger size according to the sub-implantation growth model. The amount of $sp^2$ content also increases with N concentration in the films [37, 38]. Nitrogen also acts as bridge for $sp^2$ domains and results in enlargement of $sp^2$-rich cluster size [36]. In addition, N radicals usually etch the growing film chemically and thus the film thickness decreases with N concentration as can be seen in table 1. Thus, the film density and H concentration decrease with increase in N concentration. This observation is consistent with the sub-implantation growth model. The changes in the structural properties are also reflected in the elastic modulus of the films.

Generally, the film with higher C-C $sp^3$ bonding will have higher resistance to deform in the elastic limit. In AFAM, we exactly measure the ability of the structure to deform with the help of ultrasonic wave and the coupled sample - cantilever vibration dynamics. The contact stiffness of the cantilever increases enormously for the films with higher elastic modulus which depends on the local density and the type of C-C bonding in the films. AFAM analysis confirms that the film $B_2$ has the highest modulus due to its highest density among other films. The observed elastic moduli of other films also follow a direct correlation with the density as can be seen from Fig.2 and Fig.6. Further, the measured elastic modulus is also consistent with sub-implantation growth model. Thus, we have demonstrated a direct evidence for the evolution of soft graphitic second phases in DLC films at substrate bias $\geq$ -150V. In addition, our results also confirm that nitrogen doping in DLC helps to form graphitic soft second phases at lower substrate bias.

## 5. Conclusions

Diamond-like carbon (DLC) films are grown successfully on Si (100) substrate as a function of substrate bias and nitrogen flow rate using plasma enhanced chemical vapor deposition. The evolution of graphitic structure in DLC films with increasing substrate bias and nitrogen concentration are clearly evidenced through Raman spectroscopy. The RBS and ERDA results also support Raman analysis in terms of structural transformation with the variation of density and hydrogen concentration in the films. The DLC films grown at substrate bias of -100V are found to have the highest density. Nitrogen incorporation in DLC film introduces CN molecules and it further encourages forming C-C $sp^2$ bonding. The density of film is found to decrease with increase nitrogen concentration. The hydrogen concentration also decreases with substrate bias and nitrogen concentration because of preferential hydrogen sputtering and C-N molecule formation, respectively. The elastic modulus of the DLC films is found to decrease with substrate bias and nitrogen concentration. Further, the growth mechanism of DLC films is elucidated with the help of sub-implantation model through structural evolution and elastic modulus mapping. Finally, this study provides a direct microscopic evidence for evolution of soft second phases in DLC matrix under high substrate bias and nitrogen doping through AFAM analysis.

Author information

**Corresponding author**
**\*E-mail: polaki@igcar.gov.in, kganesan@igcar.gov.in**

## Conflict of interest

The authors declare that there is no competing financial interest.

## Acknowledgements

Authors acknowledge Dr. G. Amarendra, Director, Materials Science Group, IGCAR and Dr. A. K. Bahaduri, Director, IGCAR, for their constant encouragement and support throughout the study.